\documentclass[traditabstract]{aa} 
\usepackage{txfonts,epsfig,graphicx,natbib,url,twoopt}

\newcommand{\thetabold}{\mbox{\boldmath$\theta$}}
\newcommand{\betabold}{\mbox{\boldmath$\beta$}}
\newcommand{\etabold}{\mbox{\boldmath$\eta$}}

\newcommand{\sigmabold}{\mbox{\boldmath$\sigma$}}

\usepackage{natbib,twoopt}
\usepackage[breaklinks=true]{hyperref} 
\bibpunct{(}{)}{;}{a}{}{,}             
\makeatletter
  \newcommandtwoopt{\citeads}[3][][]{\href{http://adsabs.harvard.edu/abs/#3}%
    {\def\hyper@linkstart##1##2{}%
     \let\hyper@linkend\@empty\citealp[#1][#2]{#3}}}
  \newcommandtwoopt{\citepads}[3][][]{\href{http://adsabs.harvard.edu/abs/#3}%
    {\def\hyper@linkstart##1##2{}%
     \let\hyper@linkend\@empty\citep[#1][#2]{#3}}}
  \newcommandtwoopt{\citetads}[3][][]{\href{http://adsabs.harvard.edu/abs/#3}%
    {\def\hyper@linkstart##1##2{}%
     \let\hyper@linkend\@empty\citet[#1][#2]{#3}}}
  \newcommandtwoopt{\citeyearads}[3][][]%
    {\href{http://adsabs.harvard.edu/abs/#3}
    {\def\hyper@linkstart##1##2{}%
     \let\hyper@linkend\@empty\citeyear[#1][#2]{#3}}}
\makeatother

\begin{document}

   \title{A meta-analysis of the magnetic line broadening\\in the solar atmosphere}

   \author{A. Asensio Ramos}

   \institute{Instituto de Astrof\'\i sica de Canarias,
              38205, La Laguna, Tenerife, Spain; \email{aasensio@iac.es}
            \and
Departamento de Astrof\'{\i}sica, Universidad de La Laguna, E-38205 La Laguna, Tenerife, Spain
             }
             
  \date{Received ---; accepted ---} 

  \abstract{A multi-line Bayesian analysis of the Zeeman broadening in the solar atmosphere is presented.
  A hierarchical probabilistic model, based on the simple but realistic Milne-Eddington approximation to the 
  solution of the radiative transfer equation, is used to explain the data in the optical and near infrared. Our method
  makes use of the full line profiles of a more than 500 spectral lines from 4000 \AA\ to 1.8 $\mu$m. Although the problem
  suffers from a strong degeneracy between the magnetic broadening and any other remaining broadening mechanism, the hierarchical
  model allows to isolate the magnetic contribution with reliability. We obtain the 
  cumulative distribution function for the field strength and use it to put reliable upper limits to the unresolved 
  magnetic field strength in the solar atmosphere. The field is below 160-180 G with 90\% probability.}

   \keywords{Sun: magnetic fields, atmosphere --- line: profiles --- methods: statistical, data analysis}
   \authorrunning{Asensio Ramos}
   \maketitle
%

\section{Introduction}
The physical characteristics of the outer layers of the solar atmosphere are
controlled by the magnetic field, because the magnetic pressure in the chromosphere and
corona is larger than the gas pressure. Ultimately, the magnetic energy present in these
outer layers is extracted from the energy stored in the photosphere. Consequently,
it is of interest to have reliable estimations of the magnetic energy that
is present in the photosphere.

Magnetic field diagnostics of the solar surface magnetism based on 
the polarization signals induced by the Zeeman effect are prone to
cancellations if the magnetic field is organized at scales below
the resolution element of the telescope. Despite this drawback,
our understanding of the solar (photospheric) magnetism is fundamentally based on
polarimetric studies of the Zeeman effect. Over the years, other tools
that are less or not affected by cancellations have been devised.
Among them, we can find the study of broadening mechanisms in
spectral lines \citep{stenflo77} or the analysis of the Hanle effect in
selected spectral lines \citep{stenflo82,faurobert01,trujillo_nature04}. 
It is well-known that the Zeeman signals in circular
polarization depend on the projection of the magnetic field on the line-of-sight, while in
linear polarization, they depend on the projections of the magnetic field on the plane of the sky.
In the case of the Zeeman broadening and the Hanle effect in microturbulent fields, the effect on
the spectral line depends fundamentally on the strength of the magnetic field. Therefore, no cancellation
of the effect occurs even if the magnetic field is organized at scales below the resolution element.
Both techniques have been used in the past to constrain the magnetic energy stored in the
solar photosphere. \cite{stenflo77} inferred, using the Zeeman broadening, an upper limit of 140 G for the rms
magnetic field, which would correspond to a magnetic energy of $B^2/8\pi \sim 780$ erg~cm$^{-3}$.
\cite{trujillo_nature04} obtained, using the Hanle effect in the Sr \textsc{i} line at 4607 \AA\ in the
microturbulent regime, an average magnetic field of 130 G if the intensity follows an 
exponential distribution. They also calculated that the intensity is 60 G if the field 
is homogeneous and volume-filling. This work was
extended to a realistic three-dimensional magneto-convection simulation by \cite{shchukina_trujillo11}. They
found that the field distribution in the simulation has to be increased by a factor $\sim$10 in order to
fit the observations. The ensuing magnetic energy is similar to that obtained by \cite{trujillo_nature04}
using ad-hoc distributions.

The main disadvantage of the Zeeman broadening to measure magnetic fields is that
isolating the magnetic broadening from the remaining broadening mechanisms is extremely 
difficult and uncertain. This interplay has been recently analyzed in three-dimensional simulations
of solar magneto-convection \citep{fabbian10}. In an effort to overcome the difficulties,
\cite{stenflo77} proposed to pursue a statistical approach, using Fourier Transform Spectrometer observations of many Fe \textsc{i} lines
to infer an upper limit to the magnetic field in the solar atmosphere. The idea
is based on the fact that, although the Zeeman broadening changes from line to line, it is
proportional to a factor that depends on the quantum numbers and Land\'e factors of the levels involved 
in the transition. To exploit this property, \cite{stenflo77} proposed a phenomenological formula to explain 
the measured broadening of the lines in terms of the strength of the line, excitation potential and magnetic field.
Later, \cite{solanki_stenflo84,solanki_stenflo85} exploited the same phenomenological approach to study the magnetism
of solar magnetic flux tubes, while \cite{mathys_stenflo86} and \cite{mathys89} extended this approach to other magnetic stars
with success. 

Our aim in this paper is to drop some of the previous simplifications present in the previous studies
of the Zeeman broadening. First, instead of summarizing the width of the spectral line by a single number, we
utilize the full spectral profile. Second, we use a simple but realistic line formation theory to generate
the full line profile separating the Zeeman broadening from the remaining broadening mechanisms. Third, we
analyze data in the near infrared, where the ratio between the Zeeman splitting and the Doppler broadening is
larger. Finally, we use a hierarchical Bayesian model to put reliable constraints to the magnetic field.

\section{Hierarchical modeling of line broadening}

\begin{figure*}
\centering
\includegraphics[width=\textwidth]{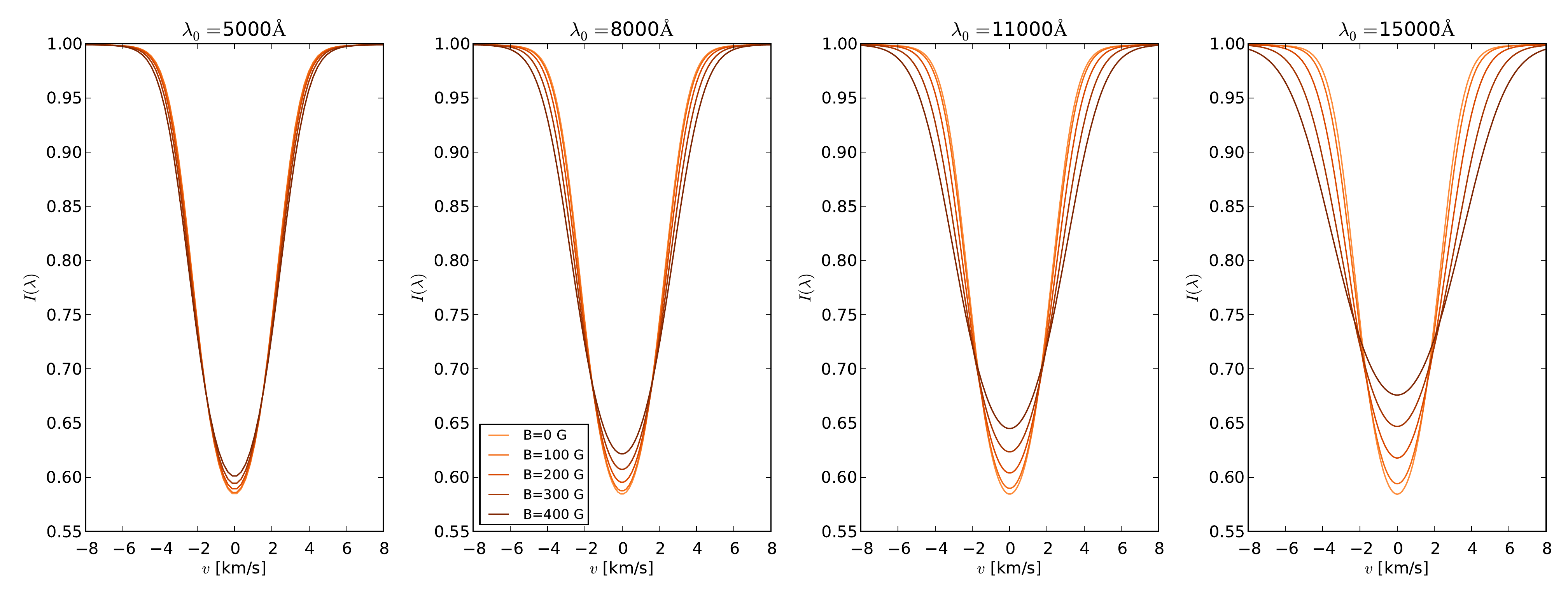}
\caption{Line profiles emerging from a Milne-Eddington atmosphere with $\beta_0=2$, $\eta_l=3$ and $a=0.02$. The lines have
a characteristic Doppler broadening of 2 km s$^{-1}$. Because the Doppler broadening is defined in velocity units, the broadening
in wavelength units increases in proportion to the central wavelength of the line. We display the line profiles for five
different microturbulent magnetic field strengths. We display the line profiles for five different microturbulent
magnetic field strengths ($B=0,100,200,300,$ and $400$~G).}
\label{fig:profiles}
\end{figure*}

\subsection{Observations}
\label{sec:observations}
In this work we use observations from the optical and the near infrared part of the spectrum, both extracted from the
National Solar Observatory (NSO) Fourier Transform Spectrometer (FTS). The observations in the optical have been
obtained from \cite{wallace_atlasvis98}, while the data for the near infrared has been extracted from \cite{wallace_atlasir03}.
From the observed spectrum, we have isolated the spectral profiles of all the Fe \textsc{i} lines tabulated 
by \cite{stenflo77} in the optical and from \cite{ramsauer95} in the near infrared.
We built a specific graphical tool to inspect the FTS spectrum and decide which lines
should be included in the study. The information from \cite{stenflo77} and \cite{ramsauer95} is
confronted with the spectroscopic information provided by the tabulation of \cite{kurucz_atomic93} to find out
if each individual line is isolated and free of blends. Lines with
clear blends in one of the wings have been also accepted but removing the wavelengths of the blend. Heavy blended
lines or those with very shallow profiles are discarded. For each line, the tool allows us to manually mark the wavelength span, which we
select as that where the intensity reaches the continuum intensity. To avoid problems in the continuum, the intensity spectrum is re-normalized to the
continuum for each line. Correlating the linelist of \cite{stenflo77} and \cite{ramsauer95} with data tabulated by
\cite{kurucz_atomic93}, we extract the central wavelength of the line, the total angular momenta
and Land\'e factors of the upper and lower level of the transition. For almost all the tabulated lines, we can
reliably find a line in the tabulation of \cite{kurucz_atomic93} with all the atomic information. 
In any of the few cases in which this does not happen, we discard the line. The central
wavelength of the line is re-computed to avoid introducing additional nuisance parameters in the
model that we describe in the following. After all the filtering, we end up with 387 spectral lines (from the original
set of 402) that span from 4365 \AA\ to 6860 \AA\ and 166 (from the original set of 352 associated to Fe \textsc{i}) 
in the range from 1 $\mu$m to 1.8 $\mu$m. The fundamental reason for the reduction in the number of lines was
the presence of a clear blend and/or very shallow lines that we decide not to use.

\subsection{Generative model and likelihood}
\label{sec:generativeModel}
Our aim is to extract information about the Zeeman broadening from a large set of spectral lines. To this end,
the first ingredient that we need is a generative model, i.e., a way of linking our physical model
with the observations. In our case, we assume that the observed spectral line at wavelength $\lambda_i$ is given by:
\begin{equation}
I^\mathrm{obs}(\lambda_i) = I(\thetabold,\lambda_i) + \epsilon_i,
\end{equation}
where $I(\thetabold,\lambda_i)$ represents the synthetic spectral line described in the following, that depends on a set of
parameters $\thetabold$. Additionally, we make the assumption that the model spectral line is perturbed with $\epsilon_i$, 
that represents noise, uncertainties and defects of our proposed model. If $\epsilon_i$ is just photon noise, it is safe to
assume that it follows a Gaussian distribution with zero mean and variance $\sigma^2$. If we allow $\epsilon_i$ to
absorb other systematic effects, this distribution becomes an approximation and one should potentially take into account the covariance between
different wavelength points. This more complicated case is surely the one that we have with
the FTS observations because the spectrum is obtained as an inverse Fourier transform of visibilities, which induces 
correlations that might be of importance. However, for the sake of simplicity and the lack of a reliable estimation for the
covariance matrix, we assume that $\epsilon_i$ follows a Gaussian distribution with zero mean and variance $\sigma^2$.

In the following, we describe the simple but powerful model that we use to synthesize a spectral line in
the presence of an isotropic and microturbulent magnetic field, i.e., we make explicit the functional form
of $I(\thetabold,\lambda_i)$ and the meaning of the vector of parameters $\thetabold$.
Under the assumption of a sufficiently weak magnetic field and neglecting magneto-optical effects, the intensity emerging in a
spectral line from a Milne-Eddington atmosphere can be written using the following simple approximate expression \citep[e.g.,][]{landi_landolfi04}:
\begin{equation}
I(\lambda) = B_0 \left[ 1+ \frac{\beta_0 \mu}{1+\eta_l \Psi(\lambda-\lambda')} \right],
\end{equation}
In the previous equation, $B_0$ and $\beta_0$ describe the variation of the source function with
optical depth, so that $S=B_0 + \beta_0 \tau$. Additionally, $\mu=\cos \theta$ refers
to the cosine of the heliocentric angle, $\eta_l$ is the ratio between the line opacity and the continuum 
opacity and $\lambda'=\lambda_0(1+v/c)$ is the central wavelength of the transition under study ($\lambda_0$) shifted by
the Doppler effect due to any bulk motion with velocity $v$. The line absorption profile
is described using a Voigt profile:
\begin{equation}
\Psi(\lambda-\lambda_0) = \frac{1}{\sqrt{\pi}} H\left( \frac{\lambda-\lambda'}{\Delta \lambda_D},a \right),
\end{equation}
where $\Delta \lambda_D=\lambda_0 \Delta v_D/c$, with $c$ the speed of light, is the Doppler width
in velocity units and $a$ is the damping coefficient.

Very far from the line center, where $\Psi(\lambda-\lambda')=0$, the continuum intensity is given by $I_c = B_0(1+\beta_0\mu)$.
Therefore, if we normalize the intensity spectrum by the continuum, we get
\begin{equation}
I(\thetabold,\lambda) = \frac{1}{1+\beta_0 \mu} \left[ 1+ \frac{\beta_0 \mu}{1+\eta_l \Psi(\lambda-\lambda')} \right].
\label{eq:radiative_transfer}
\end{equation}
It is then clear that the continuum normalized emergent intensity spectrum is independent of $B_0$.

When an isotropic, microturbulent and weak magnetic field is present in the line formation region, the model for
radiative transfer is still valid if the line absorption profile is modified. If the three components of the
magnetic field vector are characterized by a Gaussian distribution with zero mean and standard deviation $B$, the line profile
that enters into Eq. (\ref{eq:radiative_transfer}) reads \citep{landi_landolfi04}:
\begin{equation}
\Psi(\lambda-\lambda') = \frac{\Delta \lambda_D}{\Delta \lambda_T} \frac{1}{\sqrt{\pi}} H\left( \frac{\lambda-\lambda'}{\Delta \lambda_T},a \frac{\Delta \lambda_D}{\Delta \lambda_T} \right).
\end{equation}
This expression is valid at second order in the Zeeman broadening ($\Delta \lambda_B^T$), defined as:
\begin{equation}
\Delta \lambda_B^T \sim 4.6686 \times 10^{-13} \lambda_0^2 B,
\end{equation}
with $B$ given in G, and $\Delta \lambda_B^T$ and $\lambda_0$ in \AA.
We note that $\Delta \lambda_B^T \ll \Delta \lambda_D$ for the magnetic fields and temperatures expected in the quiet Sun.
According to \cite{landi_landolfi04}, the broadening of the spectral line when a microturbulent field is
present is a result of the quadratic sum of the Doppler broadening and the Zeeman broadening, so that
\begin{equation}
\Delta \lambda_T = \left[ \Delta \lambda_D^2 + 4 \bar{G}_T \left( \Delta \lambda_B^T \right)^2 \right]^{1/2}.
\label{eq:totalBroadening}
\end{equation}
The factor $\bar{G}_T$ is the \emph{second-order effective Land\'e factor for turbulent fields} and has the following
expressions in terms of the total angular momenta of the upper and lower levels and their corresponding Land\'e
factors:
\begin{equation}
\bar{G}_T = \bar{g}^2 + \frac{1}{16} g_d^2 \left( 4s-d^2-4 \right),
\end{equation}
with
\begin{align}
s &= J_u(J_u+1) + J_l(J_l+1) \nonumber \\
d &= J_u(J_u+1) - J_l(J_l+1) \nonumber \\
g_d &= g_u - g_l \nonumber \\
g_s &= g_u + g_l \nonumber \\
\bar{g} &= \frac{1}{2}g_s + \frac{1}{4} g_d d.
\end{align}
All the previous ingredients allow us to synthesize a spectral line emerging from a Milne-Eddington atmosphere
with a microturbulent magnetic field using the vector of parameters $\thetabold=(\beta_0,\eta_l,\Delta v_D,a,B)$. Examples of the influence of a microturbulent magnetic field on
the line profiles are shown in Fig. \ref{fig:profiles}. Note that the influence of the Zeeman broadening
increases with wavelength, so that it is important to focus on lines in the red part of the spectrum to
constrain the influence of a magnetic field. It is conspicuous that the effect remains extremely subtle and can be 
easily confounded with any kind of non-magnetic broadening. This is the reason why a hierarchical probabilistic 
modeling is of help, as we show in this paper.

Once the generative model is proposed, it is straightforward to write down the likelihood, $\mathcal{L}$. This
sampling distribution gives the probability of obtaining our current observations given that they have been
generated with the model presented in the previous paragraphs. In our case, our generative model results in
the following likelihood for a single spectral line $i$:
\begin{align}
\mathcal{L}_i(D_i|\thetabold_i,\sigma_i) = \prod_{j=1}^{{N_\lambda}_i} \frac{1}{\sqrt{2 \pi} \sigma_i} \exp \left[ - \frac{\left( I^\mathrm{obs}(\lambda_j) - I(\thetabold_i,\lambda_j) \right)^2}{2\sigma_i^2}\right],
\end{align}
a consequence of assuming that the intensity of different wavelengths are uncorrelated. Note that
${N_\lambda}_i$ is the number of wavelength points which are used to sample line $i$. We use the compact 
representation $D_i$ to refer to the intensity spectrum of a single line extracted from the FTS atlas.
Likewise, we use $D=\{D_1,D_2,\ldots,D_N\}$ the full set of observations.
Using the same line of reasoning, the total likelihood that takes into account all the lines is given by:
\begin{equation}
\mathcal{L}(D|\thetabold,\sigmabold) = \prod_{i=1}^{N} \mathcal{L}_i(D_i|\thetabold_i,\sigma_i),
\end{equation}
where $\thetabold=\{\thetabold_1,\thetabold_2,\ldots,\thetabold_N\}$ and $\sigmabold=\{\sigma_1,\sigma_2,\ldots,\sigma_N\}$.

\subsection{Hierarchical model}
\label{sec:hierarchical}
\begin{figure}
\centering
\includegraphics[width=0.85\columnwidth]{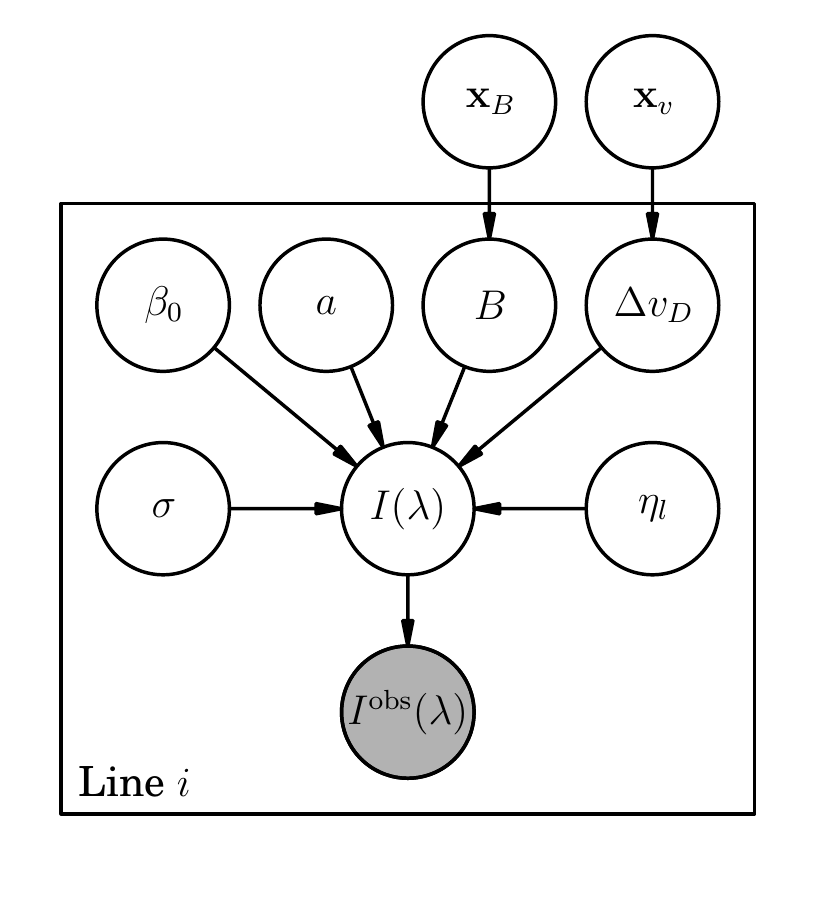}
\caption{Graphical model representing the hierarchical Bayesian scheme that we
used to analyze the broadening of spectral lines. Open circles represent random
variables (note that both model parameters and observations are considered as
random variables), while the grey circle represents a measured quantity. The frame
labeled ``Line $i$'' represents that everything inside the frame has to be repeated for all
the observations. An arrow between two nodes illustrates dependency. 
The nodes that are outside the frame are the hyperparameters of the
model and are common to all pixels.}
\label{fig:graphical_model}
\end{figure}

Taking into account all the previous considerations, the probabilistic model used in this work is graphically represented in 
Fig. \ref{fig:graphical_model}. The
intensity spectrum of each line depends on the parameters $\beta_0$, $\eta_l$, $a$, $\Delta v_D$ and $B$.
The synthetic profile, obtained applying Eq. (\ref{eq:radiative_transfer}), are then compared with the
observed profiles extracted from the FTS atlas using an unknown noise standard deviation $\sigma$. This model is repeated for all the
$N$ lines that we introduce in the analysis. 

We are interested on the statistical properties of the magnetic field when the information
encoded on all lines are simultaneously taken into account. To this end, we follow
the standard way of Bayesian hierarchical models. It consists of using parametric
priors (the parameters of the prior are termed hyperparameters) and making the inference hierarchical:
the model parameters are the lowest level while the hyperparameters represent the upper level. The graphical
representation of Fig. \ref{fig:graphical_model} gives a clear idea of the meaning of the
hierarchical model. The magnetic field strength associated to each spectral line is assumed to be extracted from a 
common distribution (prior) that
depends on the set of hyperparameters $\mathbf{x}_B$.
This hierarchical scheme can be exploited to compute the distribution of a
model parameter to which one has no direct observational access. In our case, we use a sufficiently general
parametric prior distribution and the values of the hyperparameters are
inferred from the data under the Bayesian framework. For convenience, we also use this
hierarchical structure for the Doppler velocity, making the assumption that the
Doppler widths are extracted from a common distribution.

Using the standard Bayesian approach, all the information about the model parameters is contained in the
joint posterior $p(\betabold_0,\etabold_l,\Delta \mathbf{v}_D,\mathbf{a},\mathbf{B},\mathbf{x}_B,\mathbf{x}_v|D)$, which can be computed
from the likelihood (defined above) and the prior distribution, $p(\betabold_0,\etabold_l,\Delta \mathbf{v}_D,\mathbf{a},\mathbf{B},\sigmabold,\mathbf{x}_B,\mathbf{x}_v)$, using the Bayes theorem:
\begin{align}
p(\betabold_0,\etabold_l,\Delta \mathbf{v}_D,\mathbf{a},&\mathbf{B},\sigmabold,\mathbf{x}_B,\mathbf{x}_v|D) \propto \mathcal{L}(D|\betabold_0,\etabold_l,\Delta \mathbf{v}_D,\mathbf{B},\mathbf{a},\sigmabold,\mathbf{x}_B,\mathbf{x}_v) \nonumber \\
&\times p(\betabold_0,\etabold_l,\Delta \mathbf{v}_D,\mathbf{a},\mathbf{B},\sigmabold,\mathbf{x}_B,\mathbf{x}_v).
\end{align}
Using the conditional dependencies shown in Fig. \ref{fig:graphical_model}, we can greatly simplify the likelihood and the prior distribution. The posterior for the hierarchical model becomes:
\begin{align}
p(\betabold_0,\etabold_l,&\Delta \mathbf{v}_D,\mathbf{a},\mathbf{B},\sigmabold,\mathbf{x}_B,\mathbf{x}_v|D) \propto \mathcal{L}(D|\betabold_0,\etabold_l,\Delta \mathbf{v}_D,\mathbf{B},\mathbf{a},\sigmabold) \nonumber \\
&\times p(\betabold_0) p(\etabold_l) p(\mathbf{a}) p(\sigmabold) p(\mathbf{B}|\mathbf{x}_B) p(\mathbf{x}_B) p(\Delta \mathbf{v}_D|\mathbf{x}_v) p(\mathbf{x}_v).
\end{align}
Since we are interested in the global statistical properties of the magnetic field, 
$\betabold_0$, $\etabold_l$, $\Delta \mathbf{v}_D$, $\mathbf{a}$, $\mathbf{B}$, and $\sigmabold$ can be
considered as nuisance parameters that have to be integrated out from the posterior. Therefore, our final result
is the following posterior distribution for the hyperparameters:
\begin{align}
p(\mathbf{x}_B|D) &\propto \int 
\mathrm{d}\betabold_0 \, \mathrm{d}\etabold_l \, \mathrm{d}\Delta \mathbf{v}_D \, \mathrm{d}\mathbf{a} \, \mathrm{d}\mathbf{B} \, \mathrm{d}\sigmabold \, \mathrm{d}\mathbf{x}_v
\mathcal{L}(D|\betabold_0,\etabold_l,\Delta \mathbf{v}_D,\mathbf{B},\mathbf{a},\sigmabold) \nonumber \\
&\times p(\betabold_0) p(\etabold_l) p(\Delta \mathbf{v}_D) p(\mathbf{a}) p(\sigmabold) p(\mathbf{B}|\mathbf{x}_B) p(\Delta \mathbf{v}_D|\mathbf{x}_v) p(\mathbf{x}_v) p(\mathbf{x}_B).
\label{eq:marginalization}
\end{align}

In words, our analysis proceeds as follows. We make the assumption that the shape of every line in the observations can be explained
with a parametric Milne-Eddington model that depends on the gradient of the source function, the damping coefficient, the ratio of line and continuum 
opacities and the Doppler width of the line. To this Doppler width, we add quadratically the influence of an isotropic
microturbulent magnetic field with a Gaussian distribution characterized by its standard deviation. We propose a
sufficiently general parametric distribution for this standard deviation and we use all the lines to learn something about 
these parameters. We point out that, although there are large degeneracies in the model, we will be able to learn
something about the distribution of magnetic fields thanks to the marginalization of the nuisance parameters.

\subsection{Priors for parameters and hyperparameters}
\label{sec:priors}
The prior distribution encodes all the a-priori information that we know about the parameters.
In our case, we need to put priors over the parameters $\betabold_0$, $\etabold_l$, $\Delta \mathbf{v}_D$, $\mathbf{B}$, $\mathbf{a}$, and $\sigmabold$.
The set of priors is summarized in Tab. \ref{tab:priors}, where we display the range of variation
and the type of prior. A flat prior for variable $x$ in the interval $[a,b]$ is defined as $1/(b-a)$ if $a \leq x \leq b$ and zero elsewhere. The
damping coefficient $a$, the ratio of line and continuum opacities $\eta_l$, and the Doppler width $\Delta \lambda_D$ have flat
priors. The range of variation of each parameters is chosen based on previous experience fitting the line profiles. 
Given that $\beta_0$ and $\sigma$ can potentially span over several orders of magnitude, we choose a modified Jeffreys'
prior for them \citep{gregory05}:
\begin{equation}
\mathrm{MJ}(x;x_0,x_\mathrm{max}) = \left[ \left( x + x_0 \right) \ln \left( \frac{x_0+x_\mathrm{max}}{x_0}\right) \right]^{-1}.
\end{equation}
This prior behaves as a Jeffreys' prior (i.e., as $x^{-1}$) for $x \gg x_0$ and as a 
uniform prior for $x \ll x_0$. The value of $x_0$ is chosen to be that of the lower limit, while $x_\mathrm{max}$ is
selected to be the upper limit.

\begin{figure*}
\centering
\includegraphics[width=0.24\textwidth]{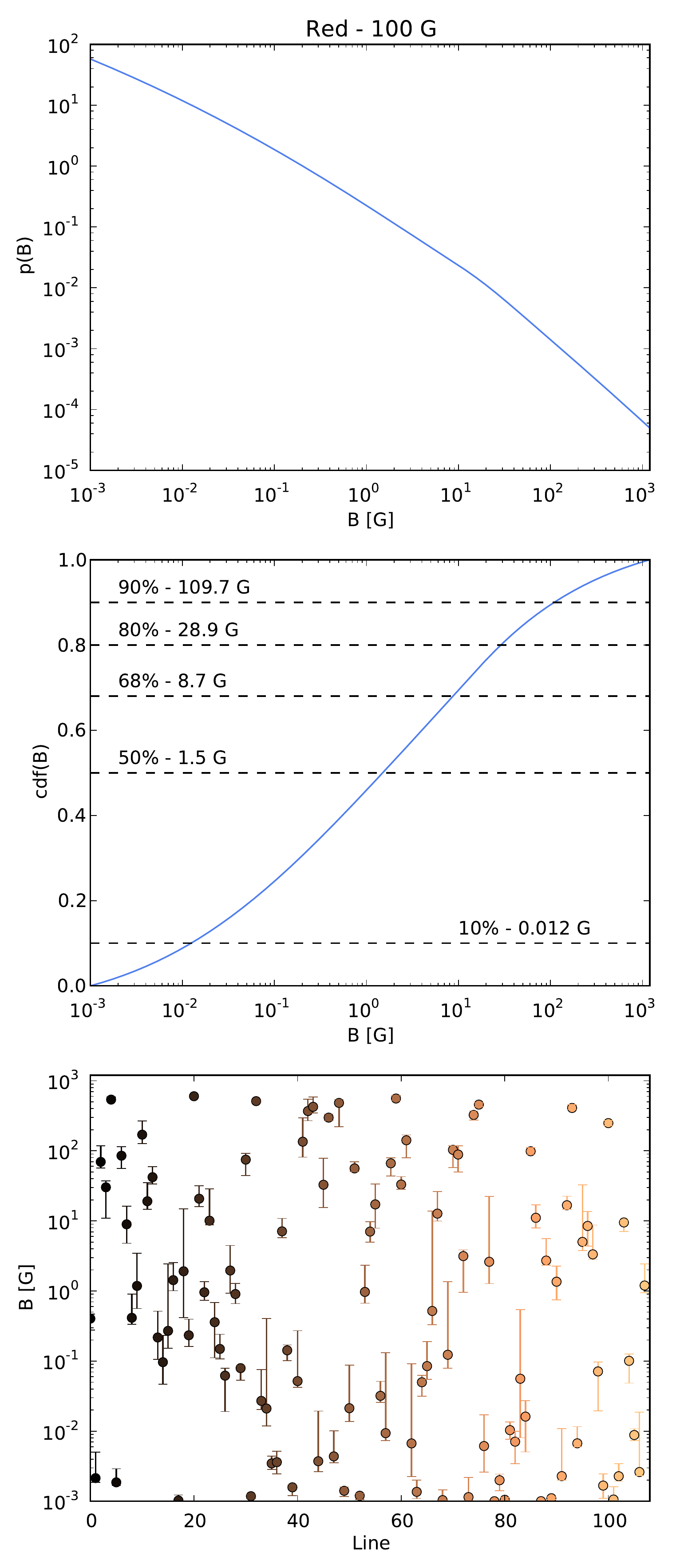}
\includegraphics[width=0.24\textwidth]{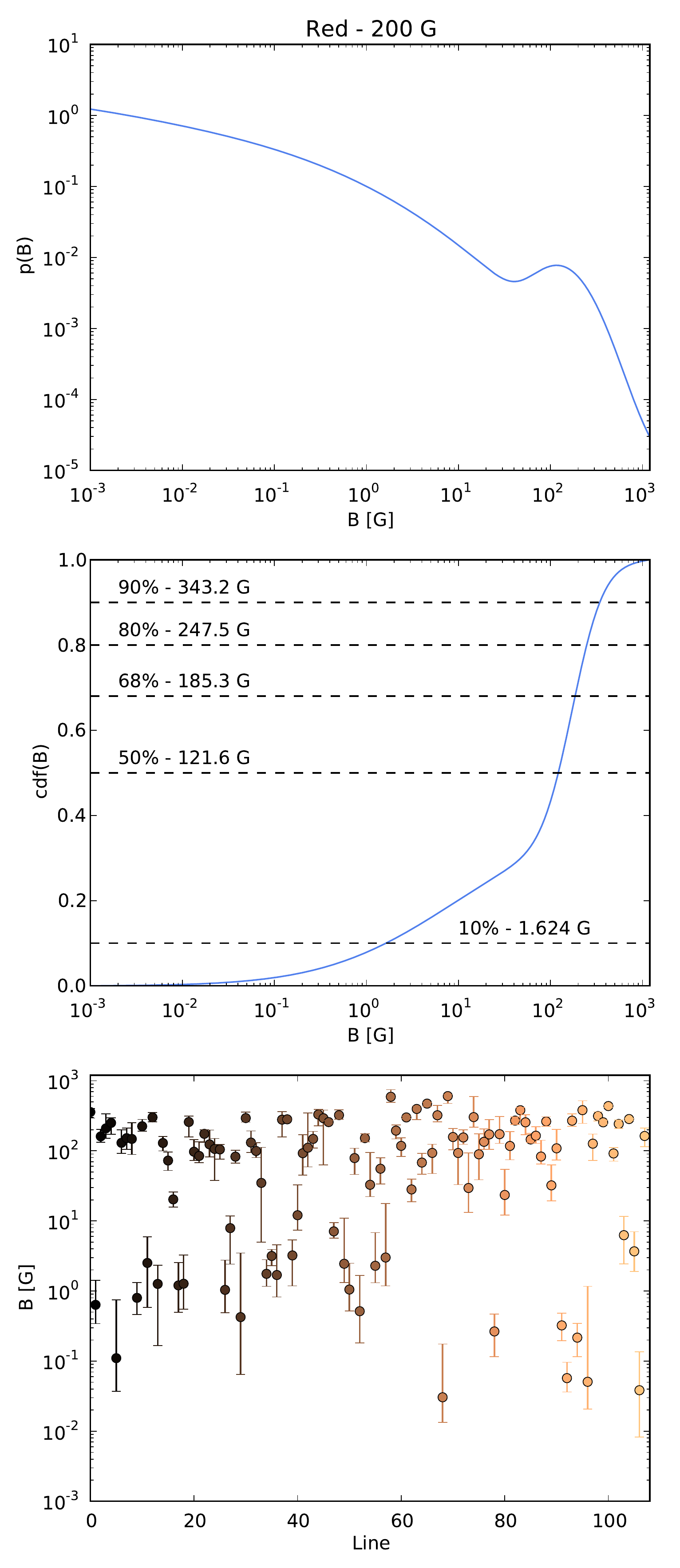}
\includegraphics[width=0.24\textwidth]{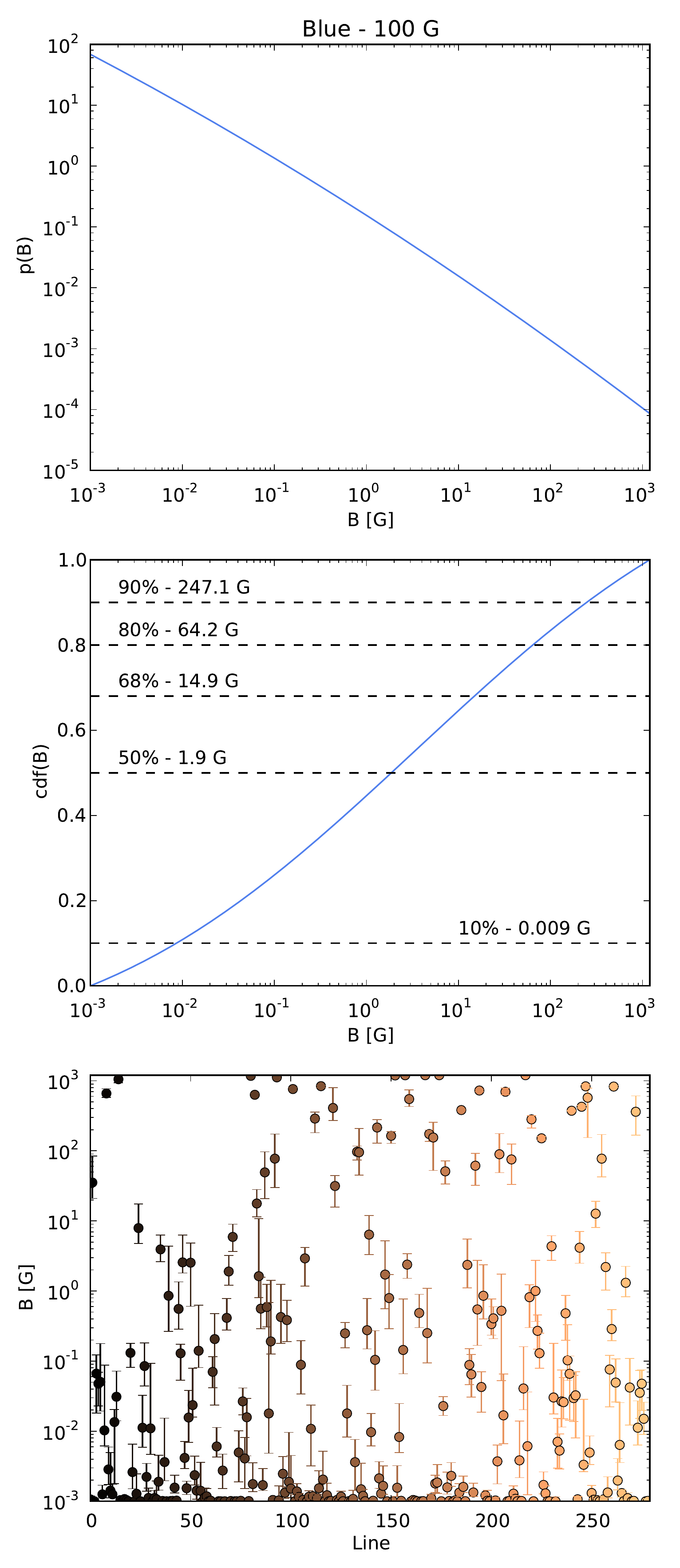}
\includegraphics[width=0.24\textwidth]{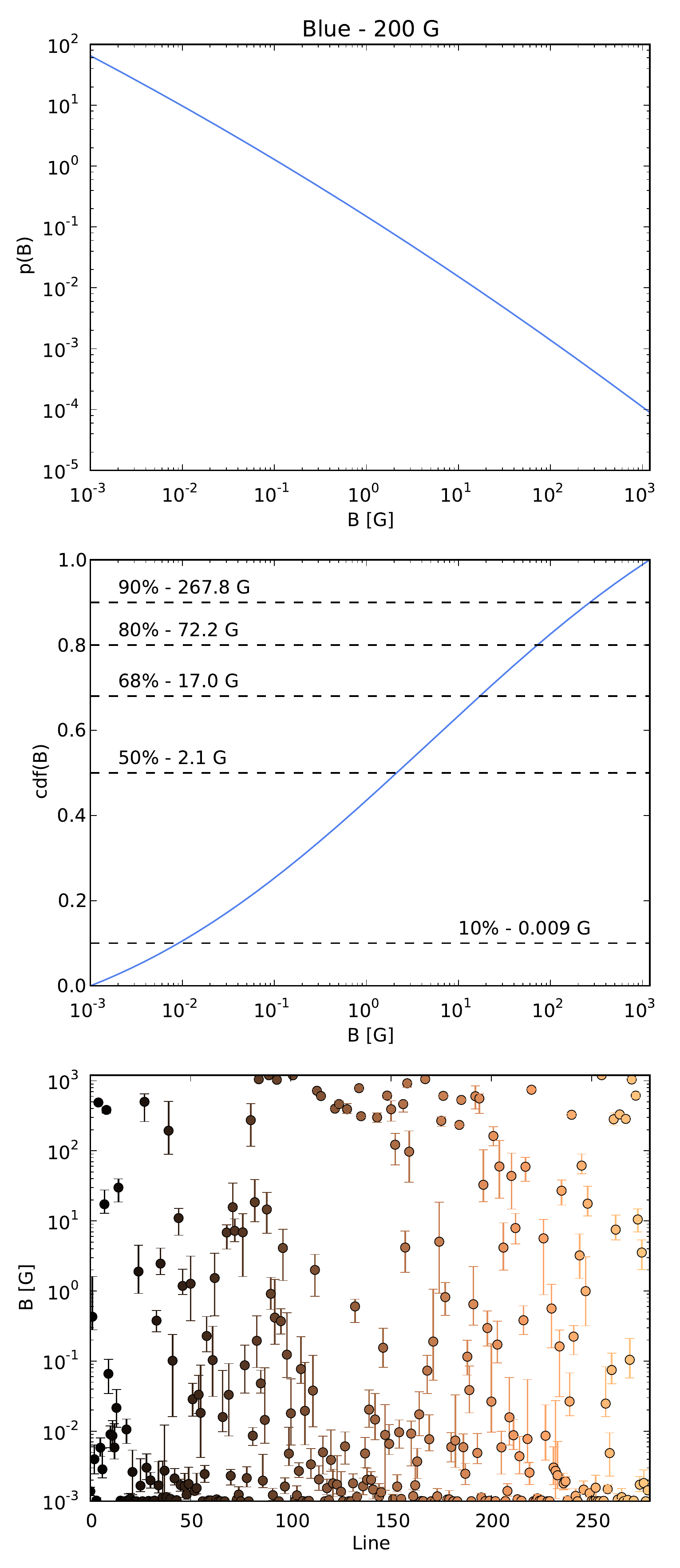}
\caption{Distribution of magnetic fields inferred from the synthetic cases. The upper panels
display the inferred probability density function. The middle panels show the corresponding cumulative distribution, 
where we have indicated the 10, 50, 68, 80, and 90\% percentiles. Finally,
the lower panels display a summary of the marginal posteriors for the field strength
associated with each spectral line. Each line has a different color in a sequence to improve the readability of the figure.}
\label{fig:hyperparameters}
\end{figure*}

Concerning the selection of the hierarchical prior distribution, it is important that the functional form is able to capture the full potential
variability of the parameter. Even with a simple prior distribution, if we allow the hyperparameters to be random variables, a wide range of
quite complex distributions can be achieved.
Additionally, a desirable property of the hierarchical prior is that it is naturally defined in the range
of the parameter of interest. A broad range of possibilities are available but we use previous successful experience \citep{asensio_arregui13} and
use simple prior distributions. After some initial experiments, we use a mixture of two log-normal prior distributions 
for $B$, truncated in the interval $[B_\mathrm{min},B_\mathrm{max}]$ (see Tab. \ref{tab:priors} for the selected values
of the limits):
\begin{equation} 
p(B_i|\mathbf{x}_B) = \left\{
\begin{array} {clc}
p \mathrm{LN}(B_i;\mu_{B1},\gamma_{B1}) & & B_\mathrm{min} \leq B \leq B_\mathrm{max} \\
+ (1-p) \mathrm{LN}(B_i;\mu_{B1},\gamma_{B2}) & & \\
0 & & \mathrm{otherwise}
\end{array}
\right.
\label{eq:prior_B}
\end{equation}
where a log-normal is defined as
\begin{equation} 
\mathrm{LN}(B_i;\mu,\gamma) = \frac{1}{\sqrt{2\pi} \gamma B_i} \exp\left[-\frac{(\log B_i - \mu)^2}{2\gamma^2} \right],
\label{eq:prior_B2}
\end{equation}
where $\mu$ and $\gamma$ are the hyperparameters, which fulfill $\gamma>0$, $-\infty < \mu < \infty$. Additionally, we have $0 \leq p \leq 1$. In the
notation used in Fig. \ref{fig:graphical_model}, we have that $\mathbf{x}_B=(\mu,\gamma)$.
One of the
main properties of this prior is that, independently of the value of $\mu$ and $\gamma$, the probability of
having $B=0$ tends to zero. This is in accordance with the fact that it is extremely improbable that the three components of 
an isotropic vector field become zero simultaneously \citep{dominguez06b,sanchezalmeida07}. 
The reason for using a mixture of two log-normals is that, in the synthetic cases displayed in Sec. \ref{sec:results_synthetic},
we find multimodal distributions for the field strength. This multimodality is a consequence of the fact
that some lines are able to measure the presence of a magnetic field, while others are quite unsensitive. Therefore,
the prior has to be flexible enough to capture this behavior. 

Finally, a single log-normal is used as prior for $\Delta v_D$, as indicated in Tab. \ref{tab:priors}. The priors for the hyperparameters and
the remaining parameters are shown in the same table. The range of parameters is chosen after running several experiments but they
can be considered to be quite ``uninformative''.

\subsection{Sampling the posterior}
For the analysis of $N=387$ lines in the optical, the posterior becomes a distribution in $6N+6=2328$ dimensions, while
it goes down to 1003 dimensions for the $N=166$ lines in the near infrared. Obviously, 
to carry out the marginalization integrals required by Eq. (\ref{eq:marginalization}) we need to rely
on standard Markov Chain Monte Carlo (MCMC) schemes \citep[e.g.,][]{metropolis53} or any other alternative. The 
dimensionality of our problem is so large that standard MCMC methods are not efficient and we resort to the
Hamiltonian Monte Carlo (HMC) method \cite{hmc_duane87,neal_hmc10}, using the code of \cite{hmc_balan13}. These methods efficiently sample the
posterior distribution by utilizing information not only of the log-posterior but also of its gradient. As a 
consequence, the sampling can make very long jumps in the space of parameters but keeping a high
acceptance rate. The main drawback is that the computation of the gradient of the log-posterior can
be cumbersome. In our case, a direct application of the chain rule allows us to compute the
gradient of the log-posterior quite efficiently. Another drawback is that HMC methods are somehow
dependent on the value of some internal parameters, that need to be adapted accordingly. We do this by an extensive
initial study in which we tweak these parameters until we get a good acceptance rate. Finally, we use a
sigmoid change of variables to improve the sampling efficiency, while avoiding sampling outside the prior ranges. 
For instance, for a parameter $x$ defined in the interval $[a,b]$, we transform it into $x'$ using
\begin{equation}
x' = \log \left( \frac{a-x}{x-b} \right),
\end{equation}
\begin{table}[!b]
\caption{Priors}
\label{tab:priors}
\centering
\begin{tabular}{cccc}
\hline\hline
Parameter & Lower limit & Upper limit & Type \\
\hline
$\beta_0$ & 0.1 & 50 & Mod. Jeffreys \\
$a$ & 0.01 & 0.6 & Mod. Jeffreys \\
$\eta_l$ & 0 & 20 & Flat \\
$\Delta v_D$ [km s$^{-1}$] & 0.2 & 6 & Log-normal \\
$\sigma$ & $10^{-4}$ & 0.1 & Mod. Jeffreys \\
$B$ [G] & 10$^{-3}$ & 1200 & Log-normal mixt. \\
$\mu_v$ [km s$^{-1}$] & 0.01 & 10 & Flat \\
$\gamma_v$ [km s$^{-1}$] & 0.01 & 20 & Mod. Jeffreys \\
$\mu_{B1}$ [G] & 0.1 & 20 & Flat \\
$\gamma_{B1}$ [G] & 0.2 & 20 & Mod. Jeffreys \\
$\mu_{B2}$ [G] & 0.1 & 10 & Flat \\
$\gamma_{B2}$ [G] & 0.05 & 10 & Mod. Jeffreys \\
$p$ & 0 & 1 & Flat
\end{tabular}
\end{table}
which is then defined in $(-\infty,\infty)$ and no boundaries are needed.
The derivatives of this transformation have to be taken into account on the derivatives of the log-posterior
with respect to the model parameters in the HMC sampling.

\begin{figure*}
\centering
\includegraphics[width=\textwidth]{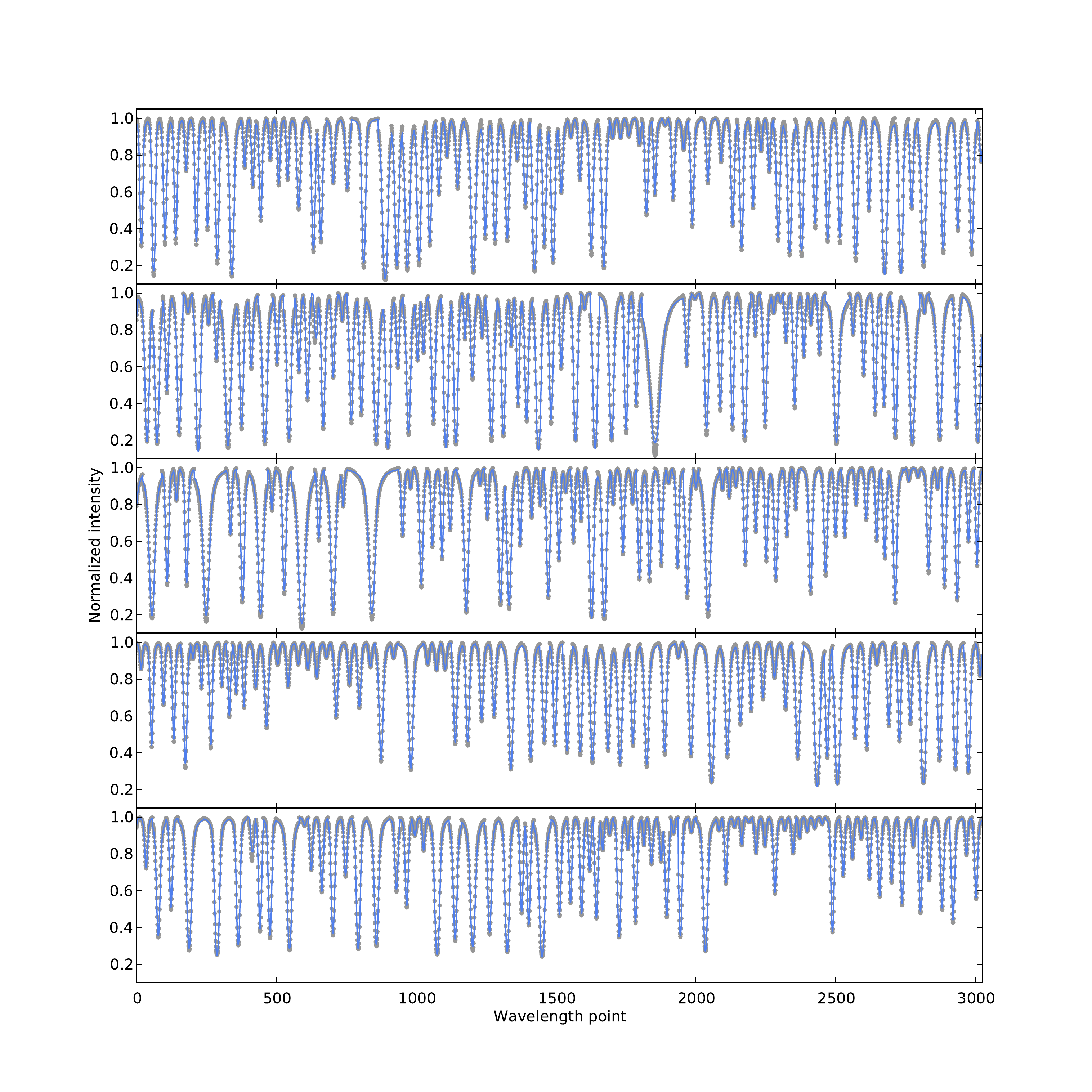}
\caption{Observed spectral lines in the optical (dots), together with the MAP fit (blue curve). The selected spectral
lines are ordered by wavelength. The abscissa axis refer to the sampling of the FTS atlas. Note that the MAP fit does
a very good job, demonstrating that the simplified Milne-Eddington model that we propose to explain the observations
is quite accurate.}
\label{fig:bestProfiles}
\end{figure*}

\begin{figure*}
\centering
\includegraphics[width=0.32\textwidth]{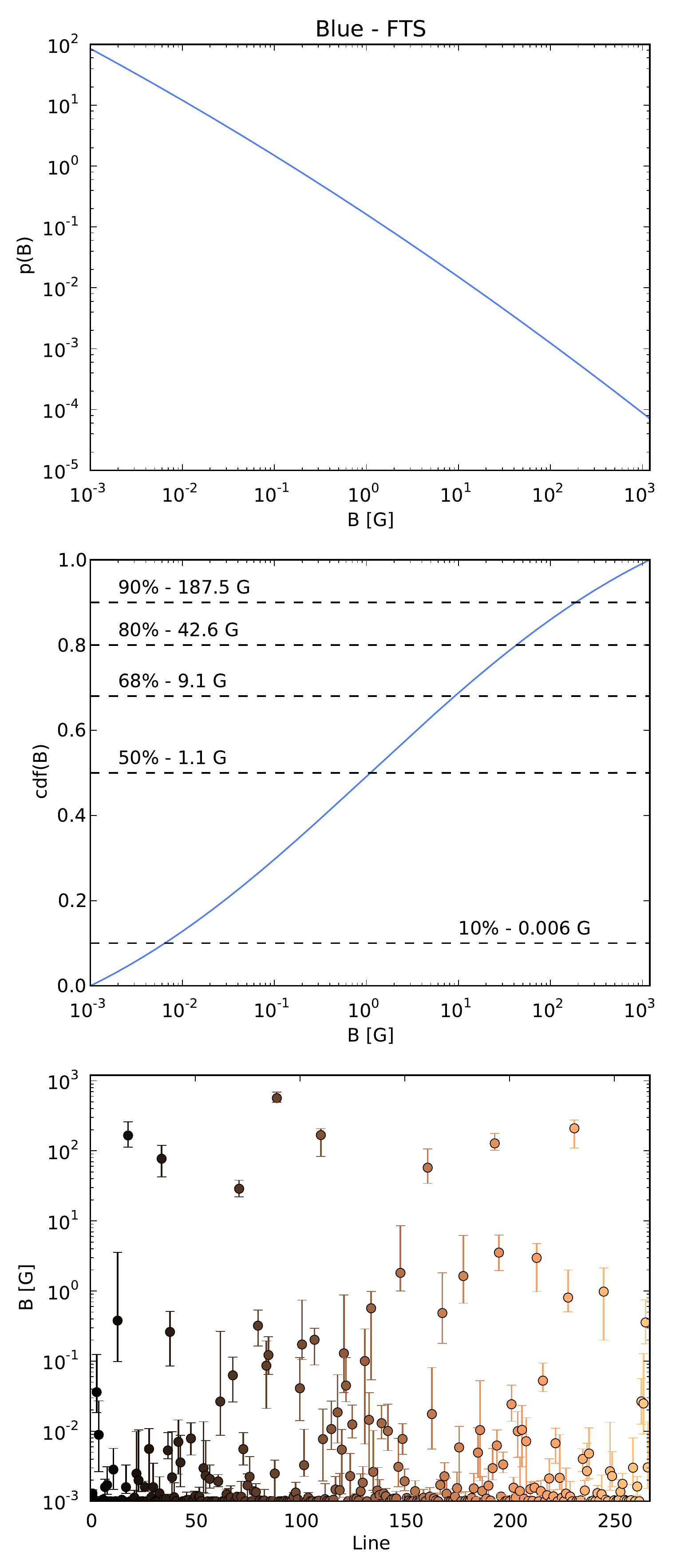}
\includegraphics[width=0.32\textwidth]{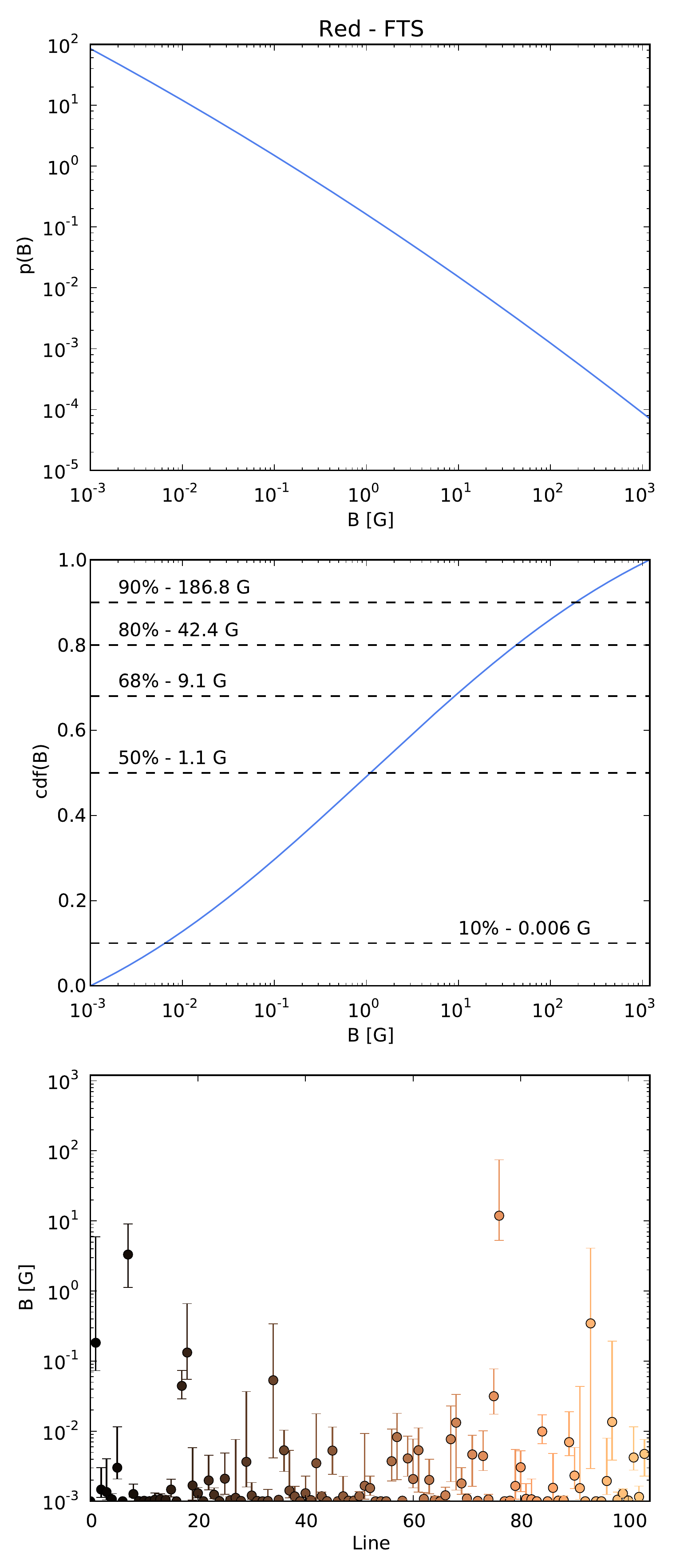}
\includegraphics[width=0.32\textwidth]{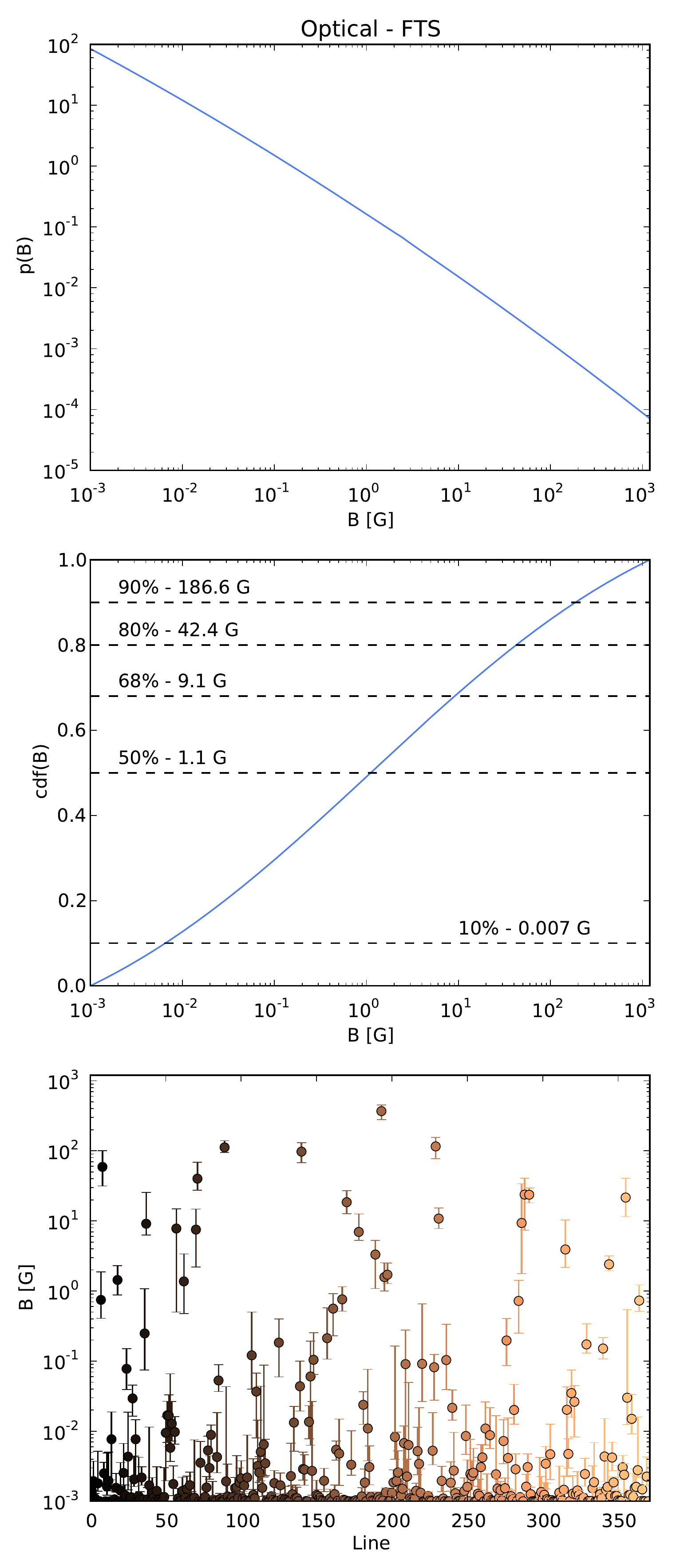}
\caption{Like Fig. \ref{fig:hyperparameters} but for the FTS atlas in the optical. For comparison, the left column displays
the results for all lines below 6000 \AA, the central column shows results for lines above 6000 \AA, while the right column
takes all lines into account. The results show that the field is below 186 G with 90\% probability.}
\label{fig:hyperparametersFTS}
\end{figure*}

\section{Results}
\label{sec:results}
\subsection{Synthetic cases}
\label{sec:results_synthetic}
The effect of a magnetic field on the broadening of a spectral line is very small and it is heavily degenerated
with other broadening mechanisms. Therefore, it is interesting to test our complex probabilistic model with a synthetic case
in which we know the input magnetic field strength.
To this end, we synthesize the 387 spectral lines using the same
Milne-Eddington model that we use to interpret it and add some noise. The line parameters are 
chosen uniformly randomly from the following intervals: $\beta_0 \in [0.5,20]$, $\eta_l \in [0.5,20]$, 
$a \in[0,0.5]$, $\Delta v_D \in [0.5,2]$ km s$^{-1}$. The magnetic field is fixed at two values, 100 G
and 200 G, and we also divide the spectral lines in two groups: lines in the blue part of the
spectrum with $\lambda_0<6000$ \AA\ and lines in the red part, with $\lambda_0>6000$ \AA.
Fig. \ref{fig:hyperparameters} display the inferred distributions for all the cases. The first
row shows the probability density function (pdf) obtained as the Monte Carlo average over the hyperparameters
of the proposed prior distribution:
\begin{equation} 
\langle p(B) \rangle = \frac{1}{N_s} \sum_{i=1}^{N_s} p_i \mathrm{LN}(B;\mu_{B1}^i,\gamma_{B1}^i) + (1-p_i) \mathrm{LN}(B;\mu_{B1}^i,\gamma_{B2}^i) ,
\label{eq:montecarlo_prior}
\end{equation}
where $N_s$ is the number of samples obtained. The second row displays the ensuing cumulative distribution, numerically
computed from the pdf. The last
row shows a summary of the inferred values of the magnetic field associated to each line. Each point marks the 
median value of the marginal posterior, while the error bars indicate the $\pm 1\sigma$ confidence intervals.
The two left columns refer to the cases using the red part of the spectrum, while the two right columns
use the lines in the blue part. It is clear from these results that the lines in the blue are not
able to put strong constraints on the magnetic field strength, even though the number of lines in the
red part is much smaller. The data in the blue is able to say that the field is below $\sim$250 G with 90\% probability
and below $\sim 60-70$ G with 68\% probability, not giving a clear hint that the two cases have different
field strengths. As shown in the upper panels, the pdf is 
fundamentally dominated by one of the log-normals (we find $p \sim 1$), so we do not detect any hint
of multi-modality. Additionally, the field inferred from
many spectral lines hit the lower boundary of 10$^{-3}$ G that we impose with the prior, giving the idea that even
weaker files would be preferred.

On the contrary, the constraints obtained from the red part of the spectrum are much more restrictive. For the case
with 200 G, the data indicates that the field is below 343 G with 90\% probability, while for the case with 100 G,
this number goes down to 109 G. The pdf for the case with 200 G gives a clear hint for the presence of a bump around $\sim 200$ G.
This bump produces that the percentile 50\% and 68\% give very similar magnetic fields (121 G and 185 G, respectively).
This experiment demonstrates that our probabilistic model is able to extract information
from the observations, in spite of the complexity and degeneracy of the problem.

\subsection{Analysis of the FTS atlas in the optical}
After the demonstration with synthetic data, we apply the very same probabilistic model to the
observed FTS atlas. Even though we have tried to isolate the clean spectral lines, the simplified Milne-Eddington 
model is surely now less appropriate than in the synthetic model to explain the spectral lines. The presence of asymmetries and/or blends
may affect our model, but they should be largely absorbed by the random variable $\sigma$. Lines where the Milne-Eddington model is
not accurate will have an enhanced value of $\sigma$. In other words, the Bayesian model automatically detects that the information
encoded in this line to constrain the model parameters is of less quality. Consequently, the relevance of these lines on the final conclusions
will be decreased.

After sampling from the posterior, we locate the combination of model parameters giving the largest posterior, i.e., 
the maximum a-posteriori (MAP) solution. The spectrum resulting from this combination of parameters
is displayed in Fig. \ref{fig:bestProfiles}. The observed spectral lines are shown in grey dots, while
the best fit is shown in with the blue curve. We only display the fitted lines for the optical. The results
in the near infrared are of the same quality. For convenience, the abscissa does not refer to real wavelength but
to wavelength step. Given the specificities of the FTS atlas, the wavelength step for the lines in the blue is 
smaller than in the red. Figure \ref{fig:bestProfiles} shows that the generative model proposed in this work
does a good job on capturing the fundamental shape of the spectral lines. The residuals are small except
in a reduced number of spectral lines.

\begin{figure}
\centering
\includegraphics[width=0.32\textwidth]{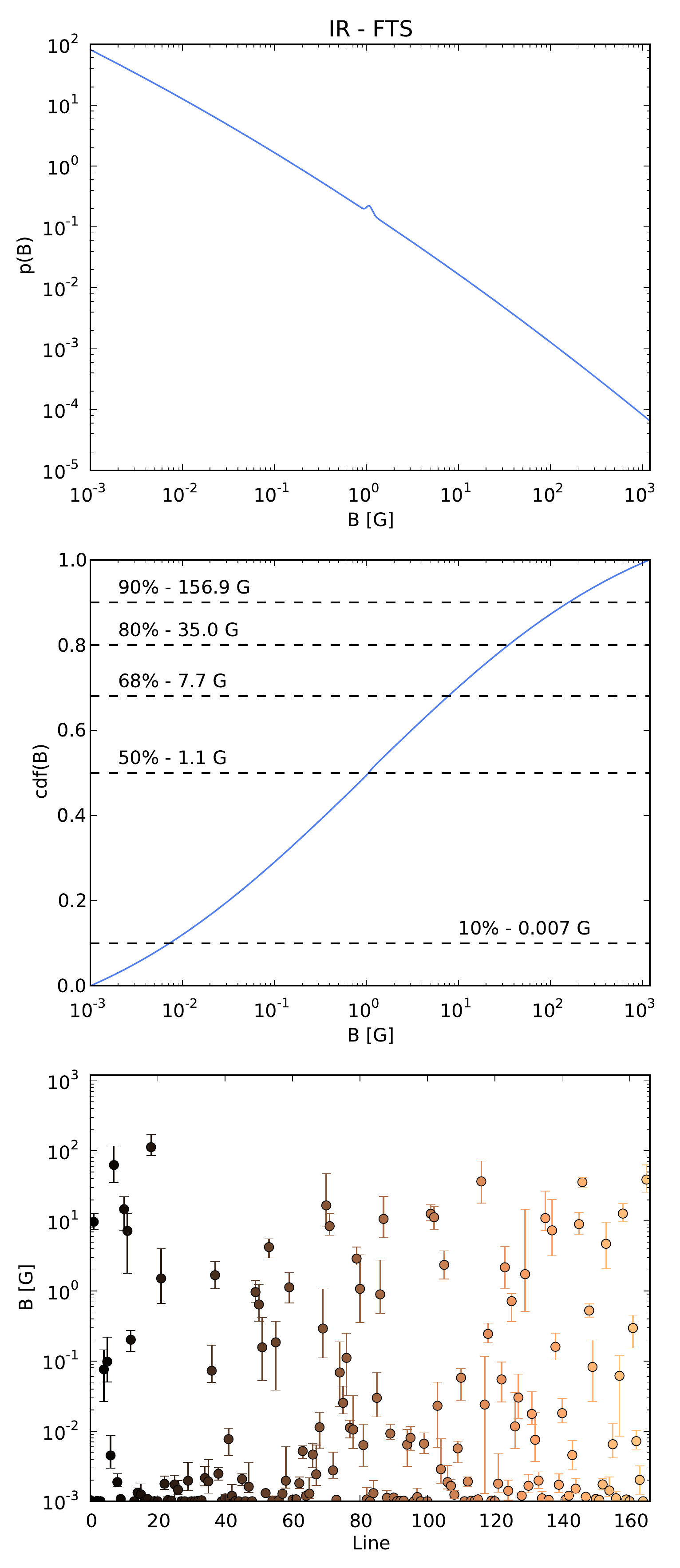}
\caption{Like Fig. \ref{fig:hyperparameters} but for the FTS atlas in the near infrared. The results in the optical
and near infrared display a good consistency, with the data in the infrared yielding fields below 160 G with 90\% probability.}
\label{fig:hyperparametersFTSVisibleIR}
\end{figure}

Having shown the quality of the generative model, we analyze now the results from the hierarchical model. 
One of the main results of this paper is displayed in Figure \ref{fig:hyperparametersFTS}. Like
in the synthetic cases, we carry out the inference separately for lines below 6000 \AA\ (blue) and above 6000 \AA\ (red),
but we also carry out the Bayesian inference for the whole set of spectra lines. Using the samples from the marginal
distribution for $\mu_{B1}$, $\gamma_{B1}$, $\mu_{B2}$, $\gamma_{B2}$ and $p$, we compute the Monte Carlo approximation
to the distribution of magnetic fields obtained from Eq. (\ref{eq:montecarlo_prior}). The corresponding 
cumulative distribution function is shown in the middle panels of the figure, where we have marked
a few quantiles of interest. The first thing to note is that the results are quite consistent independent
of the spectral range used or if we use the full set of spectral lines. Second, the probabilistic analysis demonstrates 
that the the field is, with 90\% probability, below $\sim 186$ G, although the upper limit remains hardly determined
and hits the prior limit of 1200 G. The main reason for this is that there is a degeneracy on the line broadening mechanisms, so that 
if the Doppler broadening is very small, it is still possible to fit the spectral lines using magnetic broadening. Obviously, this possibility is 
limited by the presence of the upper limit to the magnetic field set by the prior.
Finally, the analysis suggests that the majority of the spectral lines point to a very small magnetic field that
hits the lower limit of the prior. However, the influence of the specific value of the lower limit of the prior 
is of reduced importance on the general results indicated by the cumulative distribution function. Only a few lines
are the ones pushing the prior to higher magnetic fields.

\subsection{Analysis of the FTS atlas in the near infrared}
It is clear from Eq. (\ref{eq:totalBroadening}) that, for a fixed magnetic field, the Zeeman broadening is
larger in the red than in the blue. In spite of its great interest, investigating the Zeeman broadening
in the infrared has been tackled only a few times. \cite{trujillo_asensio_shchukina_spw4_06} used synthesis 
in three-dimensional simulations of the solar atmosphere with ad-hoc turbulent magnetic fields in the 
Fe \textsc{i} lines at 15648 \AA\ and 15652 \AA. They pointed out that the Zeeman broadening can be
potentially used in conjunction with other techniques to put constraints to the unresolved magnetic
field. Later, \cite{asensio_mn07} and \cite{asensio_mn09} analyzed observed and synthetic Stokes $I$ profiles of the Mn \textsc{i}
line at 15262.7 \AA\ to diagnose the magnetic field strength taking advantage of the strong Paschen-Back
perturbations produced by the hyperfine structure. They concluded that the mean magnetic field has to be smaller
than 250 G and organized at horizontal scales of $\sim$0.4''.

In this work, we apply a multiline technique to lines in the near infrared in order
to put constraints to the non-resolved magnetic field in the solar photosphere. To this end, we applied the probabilistic
model to the 166 lines that we isolated from the FTS atlas, and the results are displayed in Fig. \ref{fig:hyperparametersFTSVisibleIR}.
The cumulative distribution function points to fields below 160 G with 90\% probability, and below 35 G with 80\%
probability.

\section{Conclusions}
We have used a probabilistic model of the formation of spectral lines in a Milne-Eddington atmosphere
with a turbulent magnetic field with Maxwellian distribution to obtain the statistical properties of the average magnetic field
from the FTS atlas of the solar spectrum. The difficulty of this approach deals on the reduced effect that
a magnetic field has on the line broadening, together with the large degeneracy that the magnetic broadening
has with other broadening mechanisms. In spite of this, our hierarchical model is able to extract this hidden
information and we provide strong constraints to the turbulent magnetic field. 

The difficulty of estimating the magnetic field is reflected on the long tails that the cumulative
distribution functions have for strong fields. Even though the median value is 1 G for both the data
in the optical and near infrared (the field is below 1 G with 50\% probability), the field is smaller 
than 186 G with 90\% probability from the optical data and below 160 G with 90\% probability for the
data in the near infrared. These results are not far from the original results of \cite{stenflo77}, who obtained a value
of 140 G using the same spectral lines. Additionally, computing the magnetic energy associated to our
estimated priors, we obtain $B^2/8\pi \sim 1430$ erg~cm$^{-1}$ from the data in the optical and
$\sim 1260$ erg~cm$^{-1}$ from the near infrared data. These values are similar to those estimated
by \cite{trujillo_nature04} using the Hanle effect in the Sr \textsc{i} line at 4607 \AA.

However, our statistical analysis departs from the study of \cite{stenflo77} in a few important
details. First, we make use of the full line profile and also from data in the near infrared. We simultaneously use information from
the width of the line at all depths, contrarily to the analysis of \cite{stenflo77} that was carried out
for the width of the lines at a few discrete number of depths. Second, given our Bayesian approach
to the probabilistic model, the results are given in terms of probability distributions, from where confidence
intervals can be easily obtained. This is of special relevance in our problem, in which the desired
piece of information is deeply degenerate and hidden. It is then important to give reliable confidence
intervals. Finally, our results are given
after marginalizing all the parameters except the magnetic field. Therefore, the confidence intervals
that we extract from Fig. \ref{fig:hyperparametersFTS} already take into account our ignorance on the 
remaining model parameters. 

The magnetic field included in our model should not be strictly interpreted as microturbulent. The FTS is
an instrument with no spatial resolution inside a very large field of view, that can reach 40''. Therefore, the resulting intensity profile for each
line is the result of the addition of many profiles, each one characterized by a magnetic field.
Our assumption of a magnetic field strength with a Maxwellian distribution should then be understood 
as the global distribution of field strengths in the unresolved field-of-view.

\begin{acknowledgements}
The diagram of Fig. \ref{fig:graphical_model} has been produced with Daft (\texttt{http://daft-pgm.org}), developed by
D. Foreman-Mackey and D. W. Hogg.
We thank S. T. Balan for kindly providing the HMC code used in this paper.
Financial support by the Spanish Ministry of Economy and Competitiveness 
through projects AYA2010--18029 (Solar Magnetism and Astrophysical Spectropolarimetry) and Consolider-Ingenio 2010 CSD2009-00038 
are gratefully acknowledged. AAR also acknowledges financial support through the Ram\'on y Cajal fellowships.

\end{acknowledgements}


\end{document}